# AMIE: An annotation model for information research

ROBERT Charles Abiodun
LORIA- Campus Scientifique, B. P. 239,
54506 Vandoeuvre-Lès-Nancy, France
Tel : +33383593000
Email : abiodun-charles.robert@loria.fr

DAVID Amos
LORIA- Campus Scientifique, B. P. 239,
54506 Vandoeuvre-Lès-Nancy, France
Tel : +33383593000
Email : amos.david@loria.fr

## ABSTRACT
The objective of most users for consulting any information database, information warehouse or the internet is to resolve one problem or the other. Available online or offline annotation tools were not conceived with the objective of assisting users in their bid to resolve a decisional problem. Apart from the objective and usage of annotation tools, how these tools are conceived and classified has implication on their usage. Several criteria have been used to categorize annotation concepts. Typically annotation are conceived based on how it affect the organization of document been considered for annotation or the organization of the resulting annotation. Our approach is annotation that will assist in information research for decision making. Annotation model for information exchange (AMIE) was conceived with the objective of information sharing and reuse.

## 1. INTRODUCTION
Annotation tools are becoming very important for document interpretation. The importance of annotation as a tool in information interpretation can be seen with its popularity. Many text processors like Microsoft word, Adobe Acrobat and the like integrate features that enable users to annotate document. More than the normal use of annotation for document interpretation, annotation tool can be designed to assist in information research. We believe annotation made by different users do not only reflect the interpretation of the content of document(s) but can be used to evaluate the creator of the annotation. It is based on this basis that an annotation model AMIE is conceived.

## 2. BACKGROUND
One of the considerations in an annotation is the host document. We define document from its generalized perspective as "a trace of human activities" [14]. We can also see it as expressed concept or notion on/in a medium. We consider annotation as an action and an entity. From the perspective of an action, annotation can be defined as an act of interpretation of a document. The interpretation is of a specific context and is expressed on the document. The interpretation can be made by the producer of the document or another person. Considering it as an entity, we define it as a written, oral or graphic document usually attached to the host document.

Annotation can not take place until after the document has been available to its audience. Every annotation on incomplete document is considered as part of the initial document. Annotations will normally take a different form and look with respect to the original document. The difference in look may be noticeable in form of character used, font, style, colour or additional signs and images that is not characteristic of the original document.

## 3. EXISTING MODELS
The basic objective of annotation concept is to recuperate additional set of information that was not specified by the initial author of the document. This information is saved to the original document and referenced by a link. The goal of annotation is to allow addition to existing resources by individuals who normally will not have direct control on the original document.

We can explain most models of annotation with figure 1. A document is sent to a parser with an annotation originating from the user of the system. The parser is considered as the motor of the system. How the annotation is made and to what part of document the annotation is addressed is what makes the difference. Generally annotation is added to the document based on a specific model. An annotation with the original document is created and returned to the user of the system. This created annotation is generally in form of the original document with a link (visible or invisible) pointing to the location of resulting annotation stored in an annotation database. The location of associated annotation database is also based on several factors depending on the level of security consideration. Some annotations are stored on the application server, which demand high security considerations. This type of system enhances optimal sharing of information between users but limits privacy. Some other annotation databases are stored on local machine. In this case, it enhances higher security but limit sharing of resources. A compromise between these two types of systems is the use of proxy-storage server. In this case another machine is situated in between the application server and the local machine to store annotations.

Several annotations systems were developed along the line of the structure of the document or based on the organization of the resulting annotation [10][1]. Prominent among annotation system aimed at the structure of the document are annotation of type structuring and annotation for type classification. We can also consider annotations based on the methodology used in creating the annotations. Some are automatic, some semi-automatic and others are manual

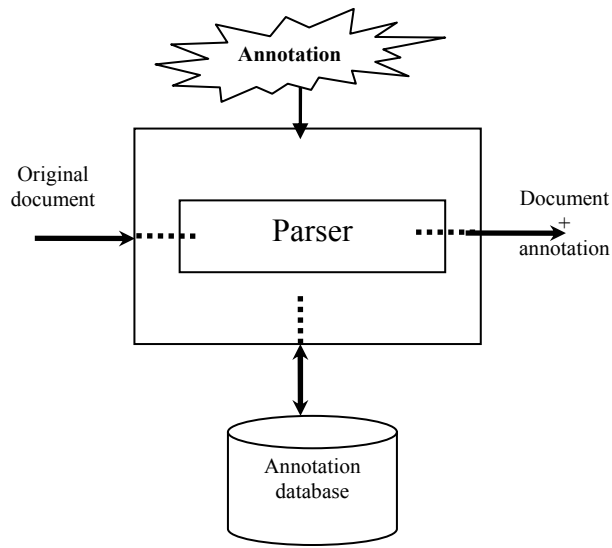

*Figure 1: Generalized annotation model*

Annotation systems based on the structure of the document are concerned with the structural relationship between resulting annotation and the elements of the document annotated. Several annotation systems on the internet were conceived along this line. Example includes "annotation engine", Hylight, AMAYA, YAWAS [8] and CritLink [19]. Annotation tools such as the one in Microsoft word is of the type structuring. Some annotation tools were conceived just to classify documents. The inside structure of documents are not addressed but the general concept or interpretation of the entire document as a whole. An example of annotation for classification is Furl (http://www.furl.net).

Annotations tools based on the methodology of creation generally give rise to semantic annotation, ontological annotation or linguistic type of annotation. Some annotations tools were considered as functional [12]. They can still be seen as either based on the organization of the resulting annotation or based on the structure of the document.

## 4. OUR APPROACH

Our approach is from the perspective of annotating document with the objective of providing a base for enhanced information research. Specifically, we want to enrich existing information database with value-added information (annotation).

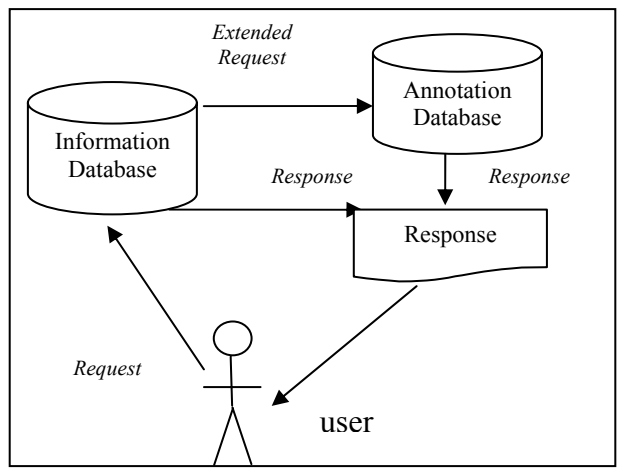

*Figure 2: Information search from annotation extension*

In a normal information research, request is sent via an interface to an information database [11]. Search terms are used to browse though the database for relevant information. Method of search varies from one system to the other. Relevant information that meets search terms is sent to the user via an interface. It should be noted that the basic principle remains the same whether the database being searched is a simple database, meta-database or multiple/integrated database.

The fact that relevant information that meets a particular search term is lost during a search may not be as a result of poor search algorithms or poor information organization. Our contemplation is that, information organization is generally based on standard criteria that may exclude vital information due to incompetence of the information organizer in the subject domain. It is therefore thought that we could extend information search beyond the underlying information database organized based on standard criteria. A search extended to added annotation database can produce excellent result. This is because added annotations are generally conceived outside the organizational protocol which gives an "out-of-norms" perspective to the underlying information.

An example: Books in the libraries are classified by their authors and other bibliographic parameters. It should be noted that these classification may not necessarily imply the view of the readers. If a book is classified as "accounting book", how can we know that this same book can be related to "medicine"? One way to know that this book can be related to other fields of studies can be through annotations that are provided by readers of the book. These readers provided annotations based on their experiences and competencies. Concretely, a reader who knew the author of a medical book may add an annotation that the author of an accounting book is a committee member of a local medical group and that his medical book can be classified as medical accounting.

## 5. AMIE MODEL

Our model consists of four main parts considered as the objects of the model: (a) the user who is also the annotator, (b) the document, (c) annotation transaction storage and (d) the process of annotation creation. We address the model from these four perspectives. Each of these objects has its characteristics and properties. We attempt to describe each.

### a. The user is the annotator
The user is identified with the following parameters.
- annotator reference (*this is a unique reference that is used to identify a user*).
- identity (*for sake of simplicity, a user is either a watcher or decision maker*)
    - His name (first name and last name)
    - Email address
    - postal address
    - region, country
    - area of activity (teaching, research, student etc)
- session (*session is used to identify his activities in the process with date and time*)

### b. Document
- document title (*original title of document*)
- descriptors and keywords (*descriptors are words used to describe the document*)
- authors (*are the producers of the document, their names and surnames*)
- date of publication of document
- format of document (PDF, word, html etc)
- abstract / résumé

### c. Annotation transaction (context of annotation stored in a storage)
This is meant to store the session of user every time the system is consulted.
- approach {the type of annotation i.e. follow up or new annotation}
- context reference (*this is a unique code for a specific annotation made by a user*)
- session reference (date/time)
- implicit parameters of the user
    - Length on system
    - Documents consulted
- explicit parameters of the users
    - user name
    - query parameters

### d. Annotation creation
- Reference (*is the reference, or code for future reference*)
- Type (*the type of annotation used*)
    » marking,
    » typographic
        ▪ italics , underlining ...
    » reformatting of text using brackets and braces,
    » passage numbering,
    » text
        ▪ in margin, footnotes, endnotes, in the gutter, by icons),
    » icons

- stars, question marks, exclamation marks,…
  - » symbols
    - to describe associations, relations between words..
- annotation location
- Why annotating (objective of annotation)?
  - » recapitulation, evaluation,
  - » summary, raise a point,
  - » classification, structuring,
  - » differentiating, for information,
  - » answer to a question,
  - » illustration, extension of document,
  - » clarify ambiguity of document

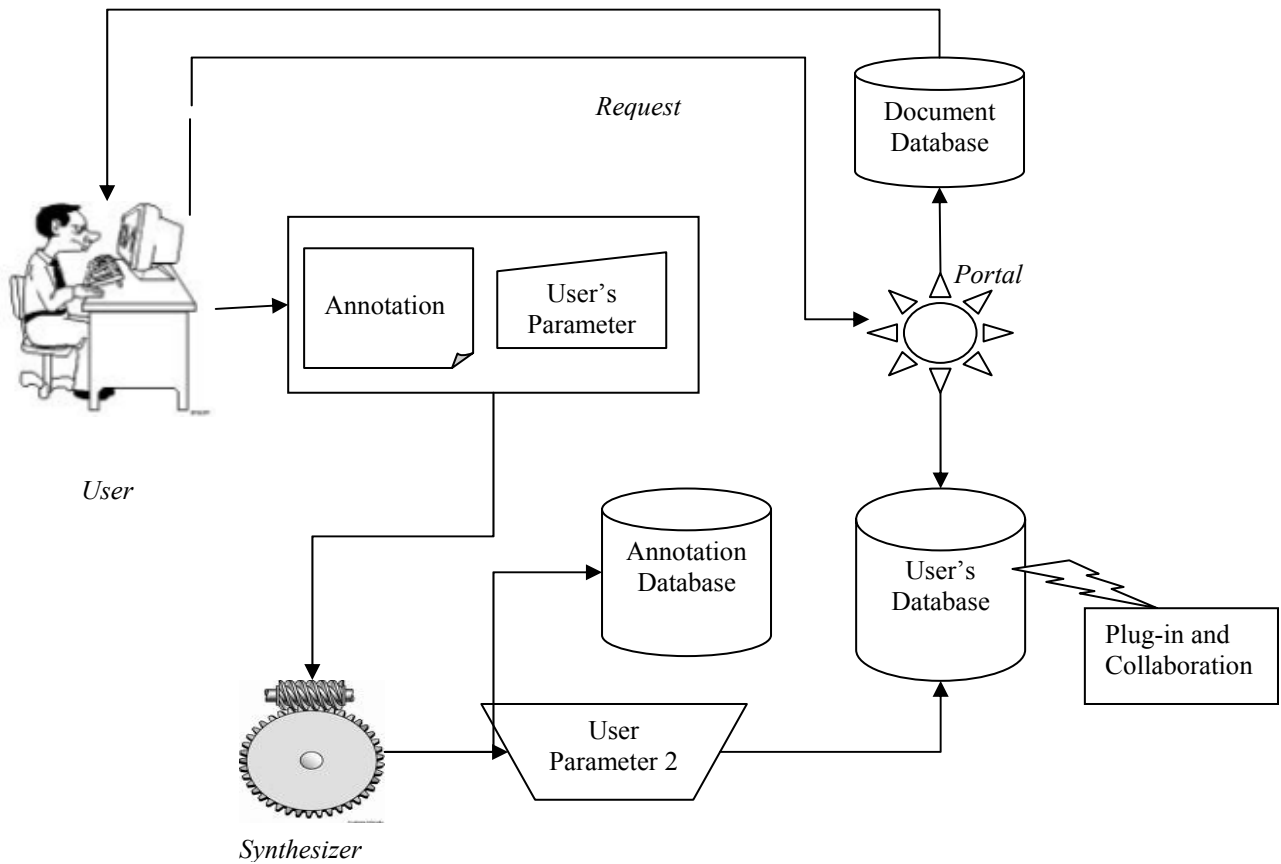

*Figure 3: Architecture of a system based on AMIE*

### Implementation

An information system based on this proposition was developed. The system can be found on http://www.loria.fr/~robert/annot. We'll describe the architecture of the system with the use of *figure 3*. A user sends his intention of using the system to the Portal. The methodology employed can be by the use of « login » and password. Just after the user access, the user formulates his request to the system in the form of « I am looking for information on a subject ». His request is sent to document database. The document database is the publications of the research SITE-LORIA. We also made provision for personal websites to be included as sources of information for annotation. We assume that the definition of the word "document" is in its generalized form so other "document" may be referenced and annotated. In a normal use of the system, documents are requested from the database. Document meeting user's request terms are retrieved from the database.

The **Portal** extracts the parameters of document(s) that was retrieved from the document database. These parameters include the file name, descriptions and document type. The **synthesizer** creates information about the user and the user's habit every time the system is uses. This is considered as the implicit profile for the user (parameter 2). Relevant document found is made available to the user who annotates the information therein. Personal information provided by the user at the beginning of system use (like the user name, address) is always made available to every user's session. This is considered as the explicit profile (parameter 1). The explicit profile with the annotations he made is sent to the users' database and annotation database respectively. The annotation is stored in the annotation database.

Any information system that wants to collaborate with this model can do so through its annotation database. Four types of collaboration were conceived to function with this system.

*A. Receptive systems.*

In this case, information is supplied to external systems: This type of system sends additional request to the annotation database of AMIE system. For example, a user may want information on "economic intelligence" from an internet site (say from www.google.fr). If response to his request is unsatisfactory, he can extend his request to the annotation database of AMIE system.

*B. Admissive systems.*

Here, the system accepts additional information from external systems: For this type of system, contribution is made to the annotation database of AMIE by external systems. A formal request is made to AMIE annotation database through its communicator before such deposit is made possible. For example, an annotation made on a bibliographic system or a financial system may be deposited in the annotation database of AMIE system.

*C. Interpretative systems.*

In the case of analyzer system, it is the system based on AMIE that makes request to external system. A result from system on AMIE is sent to external system for further clarifications or for analysis. It could also be for presentation purposes. In this case, a return is not expected from the external system. For example, the content of the databases based on AMIE can be analyzed by external system. Presently, a tool called METIORE[1] is been developed by the SITE-LORIA team to analyze the content of information stored by the system.

*D. Collaborative systems.*

This makes it possible for contributions between two systems: The system based on AMIE model is linked to another system for collaborative purpose. In this case, there are continuous exchange of information between the system based on AMIE model and another system for improvement purposes. We conceived that, an external system can be engaged in cycles of reception and transmission of information to improve the two systems involved. For example, a system based on SIMBAD model [17] can treat a certain type of information produced by AMIE. SIMBAD in this case can transform the information from our into attributes and value which can be returned to our system.

# 6. CONCLUSION

From our studies, we have made it clear that annotation can be very useful in information research. Annotation has been viewed as a function of its maker (the annotator), the document been annotated and the time of annotation. These three parameters are very important in its application to information research. We also proposed a model for annotation and how it can be interfaced with other information research and management tools. We described a system implemented based on this model.

# 7. PERSPECTIVES

The model is expected to permit analysis of stored annotation for decisional problem. For example, it should be possible to class the annotations by user group and time. It should also be possible to see at a glance the relationship between documents, relationship between users and between users and documents. The present model does not consider user age grouping. It will be good to analyze annotation base on age group because different age groups will make annotations differently. Environmental, geographical and cultural factors can be important considerations in annotation as well. It may be interesting to see the influence of technology and time on annotation.

We were not particularly concerned with ontology and semantic on annotation. It may make meaning to make annotation for information research based on ontology and semantic.

---

[1] METIORE is a system been developed by SITE-LORIA research team that will assist in analysing XML files (http://metiore.loria.fr)